\documentclass[%
 aip,
 amsmath,amssymb,
 reprint,%
onecolumn]{revtex4-1}

\usepackage{graphicx}% Include figure files
\usepackage{dcolumn}% Align table columns on decimal point
\usepackage{bm}% bold math
\usepackage[utf8]{inputenc}
\usepackage[T1]{fontenc}
\usepackage{mathptmx}

\usepackage{amsmath}
\usepackage{graphicx}
\usepackage{dcolumn}
\usepackage{amssymb}
\usepackage{color}
\usepackage{mathtools}

\usepackage{epstopdf}
\usepackage{bm}% bold math
\usepackage{enumerate}
\usepackage{fancyhdr,color}
\usepackage{verbatim}
\usepackage{amsmath,amsthm}
\usepackage{amsfonts}
\usepackage{hyperref}
\usepackage{qcircuit}
\usepackage[english]{babel}

\def\Tr{\mathrm{Tr}}
\def\Pr{\mathrm{Pr}}

\newtheorem{theorem}{Theorem}
\newtheorem{lemma}{Lemma}

\def\fig#1{Fig.~\ref{fig:#1}}

\newcommand{\ket}[1]{\left| #1\right\rangle}        % ket vector

\newcommand{\bra}[1]{\left\langle #1\right|}        % bra vector
\newcommand{\braket}[2]{\langle #1 | #2 \rangle} % <x|y>
\newcommand{\ketbra}[2]{| #1 \rangle \! \mspace{2mu}\langle #2| }
\newenvironment{proofof}[1]{\begin{trivlist}\item[]{\flushleft\it
Proof of~#1.}}
{\qed\end{trivlist}}

\begin{document}

\preprint{}

\title[]{A Variational Quantum Algorithm\\ for Preparing Quantum Gibbs States}
% Force line breaks with \\

\author{Anirban N. Chowdhury}
\email{anirban.ch.narayan.chowdhury@usherbrooke.ca}
\affiliation{D\'epartement de Physique \& Institut Quantique, Universit\'e de Sherbrooke, Qu\'ebec, Canada}

\author{Guang Hao Low}
\affiliation{Microsoft Quantum, Redmond, WA, USA}%

\author{Nathan Wiebe}
\affiliation{Microsoft Quantum, Redmond, WA, USA}%
\affiliation{Department of Physics, University of Washington, Seattle, WA, USA%\\This line break forced with \textbackslash\textbackslash
}
\affiliation{Pacific Northwest National Laboratory, Richland, WA, USA}%

\date{\today}% It is always \today, today,

\begin{abstract}
Preparation of Gibbs distributions is an important task for quantum computation.  It is a necessary first step in some types of quantum simulations and further is essential for quantum algorithms such as quantum Boltzmann training.  Despite this, most methods for preparing thermal states are impractical to implement on near-term quantum computers because of the memory overheads required.  Here we present a variational approach to preparing Gibbs states that is based on minimizing the free energy of a quantum system.  The key insight that makes this practical is the use of Fourier series approximations to the logarithm that allows the entropy component of the free-energy to be estimated through a sequence of simpler measurements that can be combined together using classical post processing.  We further show that this approach is efficient for generating high-temperature Gibbs states, within constant error, if the initial guess for the variational parameters for the programmable quantum circuit are sufficiently close to a global optima.  Finally, we examine the procedure numerically and show the viability of our approach for five-qubit Hamiltonians using Trotterized adiabatic state preparation as an ansatz.
\end{abstract}
\maketitle

\section{Introduction}
Quantum state preparation is perhaps one of the greatest outstanding challenges in quantum computing applications. In quantum simulations, preparing ground states of Hamiltonians is key to understanding problems in many-body physics and chemistry~\cite{Yung2013IntroductionTQ}. Quantum algorithms for database search, linear algebra and machine learning require preparing a quantum superposition state over input data~\cite{Grover1996search,Harrow2009linear}. Of significant importance is the problem of sampling from the Gibbs or thermal states of Hamiltonians, a problem that has wide-ranging applications such as simulating equilibrium physics~\cite{Huang1987statmech}, solving optimization problems by quantum simulated annealing~\cite{somma2008quantumSA} or quantum semi-definite programming~\cite{Brando2017QuantumSDP}, and training of Boltzmann machines in quantum machine learning~\cite{kieferova2017tomography,Biamonte2017QuantumML}. 

% State preparation is perhaps the biggest outstanding problem facing areas of quantum simulation and machine learning. In quantum simulation, preparing low energy states of a Hamiltonian is key to understanding problems such as catalysis in chemistry.  Similarly, in machine learning and optimization quantum algorithms such as Boltzmann machine training or SDP solvers hinge on the ability to prepare thermal states inexpensively~\cite{temme2011quantum,wiebe2016quantum,kieferova2017tomography,brandao2017quantum,amin2018quantum,wiebe2019generative}.  

State preparation is also a hard problem, even for quantum computers. Finding ground states of Hamiltonians is known to be QMA-hard, and therefore unlikely to admit efficient algorithms in general~\cite{Watrous2009quantumCC}. 
Under plausible complexity theoretic assumptions, preparing quantum Gibbs states at arbitrarily low temperature is as difficult as finding the ground state~\cite{Aharonov2013QuantumPCP}. 

A number of methods have nevertheless been developed to sample from thermal states on quantum computers. The most well-known approaches involve the use of quantum rejection sampling from an infinite temperature state~\cite{poulin2009sampling,wiebe2016quantum}, mixing via use of a quantum walk~\cite{yung2012quantum} or through a dynamical simulation of a system-bath interaction~\cite{riera2012thermalization,kaplan2017ground,motta2019quantum}. Even though these algorithms are expected to require an exponential amount of time in the worst case, they may be efficient if, for instance, the ratio of the partition functions of the infinite temperature state and the Gibbs state is at most polynomially large~\cite{poulin2009sampling} or the gap of the Markov chain that describes the quantum walk is only polynomially small~\cite{yung2012quantum,vanapeldoorn2017sdp}. Notably, neither method can use prior knowledge about the Gibbs state to provide an advantage for state preparation process (barring the notable exception of classical Gibbs states~\cite{wiebe2016quantum}). Further, many of these algorithms require the use of complex routines such as quantum phase estimation, which are difficult to implement on near-term devices.

In this article, we provide a new variational method to sample from the Gibbs state of a quantum Hamiltonian. Our work builds upon and generalizes the variational quantum eigensolver (VQE)~\cite{peruzzo2014VQE}, which has emerged as a popular approach for quantum simulation particularly on near-term devices~\cite{Parrish2019electronicvariational,Higgott2019variationalquantum,LaRose2018VariationalQS}. The idea in VQE is to variationally minimize the average energy as a cost function, over a suitably parametrized family of `ansatz' states. The numerical optimization is done classically, while a quantum computer is used to evaluate the energy. The use of a quantum device enables evaluating the cost function for Hamiltonians and ansatzes that would be intractable for classical computers. At the same time, the use of classical optimization reduces the amount of quantum resources needed. This makes VQE an attractive candidate for implementation on non fault-tolerant hardware, with potential applications in domains beyond quantum simulations~\cite{McClean2016theoryvariational,BravoPrieto2019variationallinear,Huang2019neartermlinear,Moll2018optimizationvariational}.

The Gibb state for a quantum Hamiltonian $H$ is defined as the density operator $e^{-\beta H}$. This state has a natural variational property that motivates our algorithm\,---\,it is the state that minimizes the free energy of a system at a fixed temperature~\cite{reif1998fundamentals}. We recall that the free energy of a ssytem described by a density operator $\sigma$ is given by $F= \Tr{(\sigma H)}-kT\Tr{(\sigma\log\sigma)}$ where $T=[k\beta]^{-1}$ is the temperature of the system. Therefore, if one has a circuit capable of producing a family of density operators that are parameterized by the vector $\vec{\theta}$ then the Gibbs state  $\rho$ is approximately
$$
\rho\approx {\rm argmin}_{\vec{\theta}}\left( {\rm Tr}(\sigma(\vec{\theta}) H) - kT{\rm Tr}(\sigma(\vec{\theta}) \log \sigma(\vec{\theta}))\right)\;.
$$
Designing a variational algorithm based on the above requires an efficient procedure for evaluating the free energy on a quantum computer. This is one of the main questions that we address in this paper. A key challenge proves to be the estimation of the von Neumann entropy, $S(\sigma) = -\Tr(\sigma\log\sigma)$.  
We propose a procedure for estimating the free energy that uses tools such as quantum amplitude estimation along with recently-developed techniques for approximating operators by a linear combination of unitaries. We show that the von Neumann entropy can be approximated as a finite sun of traces of unitary operators. These unitaries in fact correspond to $\exp(-i\sigma t_m)$ for different values of $t_m$ and can be implemented by algorithms for density matrix exponentiation~\cite{Lloyd2014QuantumPC,lowchuang2016qubitization}. The trace estimation is then done via iterative quantum phase estimation~\cite{kitaev1996quantummeasurements}.

Our hybrid quantum-classical algorithm aims to prepare an approximation to a purification of the thermal state for a fixed Hamiltonian. To provide an estimates of the complexity of our algorithm, we consider using gradient descent to minimize the free energy, and prove an upper bound on the complexity of estimating gradients of the free energy. 

The efficiency of our approach depends on the ability to use intuition about the form of the Gibbs state to prepare an ansatz state that can be moulded through a hybrid quantum-classical optimization method into the Gibbs state. Thus it succeeds and fails for different reasons than existing methods and so is of independent value to those approaches.

This paper is laid out as follows. First, in Section~\ref{sec:free_energy_estimation} we explain our quantum algorithm for estimating the free energy. We begin by stating our main result, and then go into a detailed discussion of the algorithms for estimating the von Neumann entropy and the average energy. Section~\ref{sec:free_energy_optimization} concerns itself with the complexity of using gradient descent to minimize the free energy. We then transition in Section~\ref{sec:near_term} into a discussion on how these methods could be executed in near-term devices and provide numerical examples that demonstrate the validity of our approach. 

% The paper is laid out as follows. We begin in Section~\ref{sec:entropy_est} with a discussion of our algorithm for estimating the von Neumann entropy and also enumerate its complexity. This is followed by an explanation of the energy estimation in Section~\ref{sec:energy_est}. 
% First we discuss the problem of free-energy optimization and state our main result. We then discuss entropy estimation and the complexity of energy estimation and use these results to prove our main result. Next we discuss the complexity of estimating the gradient and then transition into a discussion on how these methods could be executed in near-term devices and conclude by providing numerical examples showing the efficacy of the approach.
%We provide formal estimates of the complexity of this task and give small-scale numerical experiments that illustrate our method.  

%The goal is to prepare a thermal Gibbs state of a $D$-dimensional system using a variational approach. The Gibbs state is the distribution that minimizes the Helmholtz free energy, $F = E(\rho)-TS(\rho)$. We assume that we have access to a protocol with parameters $\vec{\theta}$ which generates a purified version of the density matrix $\ket{\rho(\theta)}$. The Gibbs state will be obtained by performing gradient descent on the free energy. 

\section{Free Energy Estimation}\label{sec:free_energy_estimation}

\subsection{Entropy Estimation}\label{sec:entropy_estimation}

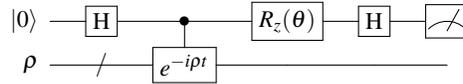
\begin{figure}
\begin{equation*}
%\Qcircuit @C=1.5em @R=0.3 em {
%               &  & \lstick{\ket{0}} & \gate{H} & \ctrl{1} & \gate{H} & \meter & \push{\rule{-1em}{4em}} \\
%               &  & & \qw & \multigate{4}{e^{-i\rho t}} & \qw & \qw \\
%               &  & & \qw & \ghost{e^{-i\rho t}} & \qw & \qw \\
%               &  &\lstick{\sum_j \sqrt{p_j}\ket{j}\ket{\chi_j}} & \qw & \ghost{e^{-i\rho t}} & \qw & \qw \\
%               &  & & \qw & \ghost{e^{-i\rho t}} & \qw & \qw \\
%               &  &  & \qw & \ghost{e^{-i\rho t}} & \qw & \qw \gategroup{2}{3}{6}{5}{0.5em}{\{}
%        }
\Qcircuit @C=1.5em @R=0.3 em {
               & \lstick{\ket{0}} & \gate{\rm H} & \ctrl{1} &\gate{R_z(\theta)} & \gate{\rm H}  & \meter  \\
               & \lstick{\rho} & \qw {/}  & \gate{e^{-i\rho t}}&\qw  & \qw & \qw 
        }   
\end{equation*}
\caption{Circuit for estimating individual terms in a Fourier series.  This circuit corresponds to the standard circuit for iterative phase estimation, with the notable exception that density matrix exponentiation is used on the input state $\rho$.  The phase $\theta$ can be chosen to be $0$ or $\pi$ which correspond to the cosine and sine components of the Fourier series respectively.} 
\label{fig:entropyestcircuit}
\end{figure}

Even though there exist quantum algorithms for estimating certain entropic quantities using a quantum computer, such as~\cite{li2018quantum,subramanian2019quantum}, these have an explicit polynomial dependence on the Hilbert-space dimension of quantum system, and are thus may not be suitable for our purposes. We propose an entropy-estimation algorithm which involves using matrix exponentiation in concert with the iterative phase estimation circuit in~\fig{entropyestcircuit} to build a Fourier series approximation to the von Neumann entropy. The overall complexity of this procedure is given below.
\begin{theorem}
\label{thm:est_entropy}
Given access to an oracle $ U_{\rho} $ that prepares a purification $ \sum_j \sqrt{p_j} \ket{j}\ket{\psi_j} $ of a density matrix $ \rho = \sum_j p_j \ketbra{\psi_j}{\psi_j} $ where $p_j \ge p_{\min}$, the von-Neumann entropy of the state $\rho$ can be estimated within error $\epsilon$ with probability greater than $2/3$ with $\widetilde{O}(1/p_{\min}^2 \epsilon)$ preparations of the purified density operator.
\end{theorem}
The behavior of the circuit used to learn the Fourier components in the expansion of the entropy is formally stated below.
\begin{lemma} \label{lem:eval_fourier_coeff}
For any $t\in \mathbb{R}$ and $\theta\in \mathbb{R}$ in the circuit of~\fig{entropyestcircuit} the probability of yielding $y \in \{+,-\}$ is\\
$\Pr(y|\theta) = \frac{1}{2}\left(1 + y {\rm Tr}( \rho \cos(\rho t + \theta/2) ) \right)$.
\end{lemma}
\begin{proof}
\begin{align}
\ketbra{0}{0} \otimes \rho &\mapsto_{H} \frac{1}{2} (\ketbra{0}{0} + \ketbra{1}{0} + \ketbra{0}{1} + \ketbra{1}{1}) \otimes \rho\nonumber\\
&\mapsto_{\Lambda(e^{-i\rho t})} \frac{1}{2} (\ketbra{0}{0}\otimes \rho + \ketbra{1}{0}\otimes \rho e^{-i\rho t} + \ketbra{0}{1} \otimes \rho e^{i\rho t} + \ketbra{1}{1}\otimes \rho)\nonumber \\
&\mapsto_{R_z} \frac{1}{2} (\ketbra{0}{0}\otimes \rho + \ketbra{1}{0}\otimes \rho e^{-i(\rho t+\theta)} + \ketbra{0}{1} \otimes \rho e^{i(\rho t+\theta)} + \ketbra{1}{1}\otimes \rho)\nonumber \\
& \mapsto_H \frac{1}{4} \left(\ketbra{0}{0} \otimes \rho (2\openone + 2\cos(\rho t +\theta) ) +\ketbra{1}{1} \otimes \rho (2\openone - 2\cos(\rho t +\theta) )\right) +\cdots
\end{align}
where above we neglect all terms that have zero trace.  Thus if we interpret the $0$ / $1$ outcomes of the measurement as $+$ / $-$ respectively then we find that
\begin{equation}
\Pr(y|\theta) = {\rm Tr} \left( \frac{1}{2}\left( \rho +  y\rho\cos(\rho t +\theta)\right) \right)= \frac{1}{2}\left(1 + y {\rm Tr}( \rho \cos(\rho t + \theta) ) \right).
\end{equation}
\end{proof}

%The core of our entropy estimation algorithm is the quantum circuit presented in Fig.\ \ref{fig:entropyestcircuit}
%The top qubit is initialized in the $\ket{+}$ state, while the others encode a purified density matrix. We assume that there is a preparatory unitary $U_{\rho}$ that prepares the latter, i.e., 
%\begin{align} \label{eqn:preparerho}
%U_{\rho}:~\ket{0}\longrightarrow \sum_j \sqrt{p_j} \ket{j}\ket{\chi_j}
%\end{align}
%The gate $R_{Z}(\theta)$ applies a rotation by either $\theta=0$ or $\pi$. Then, controlled on the top qubit, we perform a controlled unitary $U=e^(-i\rho t)$ via density matrix exponentiation. Finally we measure the top qubit in the $\ket{\pm}$ basis. The probability of measuring $\pm 1$ is given by
%\begin{align} \label{eq:probmeas}
%\text{Pr}(\pm 1) = \frac{1}{2}\pm\frac{1}{2}\sum_j p_j \cos{p_j t} \qquad \text{or} \qquad \text{Pr}(\pm 1) = \frac{1}{2}\pm\frac{1}{2}\sum_j p_j \sin{p_j t},
%\end{align}
%for $\theta=0$ or $\pi$ respectively. This proves Lemma \ref{lem:eval_fourier_coeff}
In particular, choosing $ \theta =0 $ and $ \theta=\pi $ gives us the following probabilities,
\begin{align}
\Pr(y|0) = \frac{1}{2}\left(1 + y {\rm Tr}( \rho \cos(\rho t) ) \right), \\
\Pr(y|\pi) = \frac{1}{2}\left(1 + y {\rm Tr}( \rho \sin(\rho t) ) \right).\label{eq:probmeas}
\end{align}
The implies that given a Fourier series approximation to a function $ f(x) $, we can evaluate the functional $\rho f(\rho)$ of a density matrix $\rho$ through repeated uses of the circuit in Fig.\ \ref{fig:entropyestcircuit}. 
%The von Neumann entropy is defined as, 
%\begin{align}
%S'(\rho) = -\text{Tr}(\rho\log\rho) = -\sum_{j=1}^N p_j\log p_j \;.
%\end{align}
%The thermodynamic entropy is given by $k_B/\ln 2$ times the von Neumann entropy, where $k_B$ is the Boltzmann factor.
We will therefore aim to find coefficients $ b_m^{(1)}, b_m^{(2)} $ and times of evolution $ t_m $ such that, for all $p_j\in [p_{\min},1]$, we have
\begin{equation}
\left|-\sum_{j=1}^Np_j \left[\log p_j -  \sum_m \left(b_m^{(1)} \cos (p_j t_m) + b_m^{(2)} \sin (p_j t_m) \right)\right]\right|\le \epsilon,
\end{equation}
for some $\epsilon>0$.
We achieve this by approximating the logarithm with a truncated Taylor series which is then converted to a Fourier series. As the logarithm is a divergent function, we do not expect to have a uniformly convergent series. However, our results imply that if we are bounded away from the singularities then we can construct such a series using a logarithmic number of terms. If, on the other hand, the region of interest approaches the singularity then the number of terms scales inversely with the distance from the domain of interest and the singularity. 
We begin by showing in the lemma below the maximum degree of the Taylor series required to approximate the logarithm.
\begin{lemma} \label{lem:logtaylorapprox}
For $p\in [p_{\text{min}},1]$ and all $\epsilon > 0$, $\exists~K\in\mathbb{Z}_+$ such that,
\begin{align}
\left|\ln(p)-\sum_{k=1}^{K} a_k (1-p)^k \right| \leq \frac{\epsilon}{4}
\end{align}
where 
\begin{align}
K\in\Theta\left(\frac{\log(1/\epsilon)}{p_{\min}}\right), \qquad \|a\|_1 \in O\left(\log\left(\frac{\log(1/\epsilon)}{p_{\min}}\right) \right)
\end{align}
\end{lemma}
\begin{proof}

Truncating the Taylor series of $\ln(p)$ for $p\in [p_{\text{min}},1]$ at order $K$ gives,

\begin{align}
\left|\ln (p)-\sum_{k=1}^K \frac{(1-p)^k}{k} \right| &= \sum_{k=K+1}^{\infty} \frac{(1-p)^k}{k} \leq \frac{1}{K+1} \sum_{k=K+1}^{\infty}(1-p)^k \leq \frac{1}{K+1}\int_K^\infty (1-p)^k \mathrm{d}k \nonumber\\
&= \frac{(1-{p_{\min}})^{K}}{(K+1){\log(1/p_{\min})}}\le \frac{(1-{p_{\min}})^{K}}{K+1}.
\end{align}
This error is at most $\epsilon$ if
\begin{equation}
K \ge -1 +\frac{W_0\left(\frac{-\log(1-p_{\min})}{\epsilon(1-p_{\min})} \right)}{-\log(1-p_{\min})}
\end{equation}
To suppress the r.h.s. to $O(\epsilon)$, it suffices to choose 
\begin{align}
K \in \Theta\left(\frac{\log(1/\epsilon(1-p_{\min}))}{\log(1/[1-p_{\min}])}\right)=\Theta\left(\frac{\log(1/\epsilon)}{p_{\min}}\right) \label{eq:taylortrunc}
\end{align}
the last step being valid for ${p_{\min}}\in (0,1]$.
Also, the coefficients in the expansion add up to $\|a\|_1=\sum_{k=1}^K \frac{1}{k} \leq \ln K+1$ and hence
\begin{align}
\|a\|_1 \in O(\log K)\;. \label{eq:taylornorm}
\end{align}
The proof follows from \eqref{eq:taylortrunc} and \eqref{eq:taylornorm}.
\end{proof}

Next we appeal to a lemma by van Apeldoorn et al. \cite{vanapeldoorn2017sdp} that allows us to efficiently convert the truncated Taylor series to a Fourier series with degree that is almost the same (up to logarithmic factors). We state their result below for completeness.

\begin{lemma}\cite{vanapeldoorn2017sdp} \label{lem:taylortofourier}
	For $\delta \in (0,1)$ and $f:\mathbb{R}\rightarrow \mathbb{C}$ s.\ t.\ $\left|f(x)-\sum_{k=0}^Ka_kx^k \right|\leq \epsilon/4$ for all $x\in [-1+\delta,1-\delta]$ and $\epsilon \in (0, 4\|a\|_1]$. Then $\exists c \in \mathbb{C}^{2M+1}$ s.\ t.\
	\begin{align}
	\left|f(x)-\sum_{m=-M}^M c_me^{i\pi mx/2 } \right| \leq \epsilon
	\end{align}
	for all $x\in[-1+\delta, 1-\delta]$,  where $M=2\left\lceil\ln\left(\frac{4\|a\|_1}{\epsilon}\right)\frac{1}{\delta} \right\rceil$ and $\|c\|_1\leq \|a\|_1$. Moreover $c$ can be efficiently calculated on a classical computer in time $\text{poly}(K,M,\log(1/\epsilon))$.
\end{lemma} 
The above lemma, along with the Taylor series approximation of Lemma~\ref{lem:logtaylorapprox} leads to a Fourier series approximation for $\ln p$. 
\begin{lemma} \label{lem:logfourierapprox}
For $p\in[p_{\text{min}},1]$ and all $\epsilon >0$, $\exists~M\in\mathbb{Z}_+$ such that,
\begin{align}\label{eq:logfourierapprox}
\left|\ln p - \sum_{m=1}^{M}\left(b^{(1)}_m\cos(p t_m)+b^{(2)}_m\sin(p t_m) \right)\right| \leq \varepsilon
\end{align}
where
\begin{align}
M\in O\left(\frac{\log(1/p_{\text{min}}\epsilon)}{p_{\min}} \right), \qquad \|b\|_1=\|b^{(1)}_m\|+\|b^{(2)}_m\| = O\left(\log\left(\frac{\log(1/\epsilon)}{p_{\min}}\right) \right)
\end{align}
\end{lemma}
\begin{proof}

It follows from Lemmas \ref{lem:logtaylorapprox} and \ref{lem:taylortofourier} that we can find 
\begin{align}
f(p) = \sum_{m=-M}^M c_m e^{i\pi m (1-p)/2}
\end{align}
such that $\left|\ln(p)-f(p) \right| \leq \epsilon $ where the maximum degree is
\begin{align}
M&\in O\left(\frac{1}{{p_{\min}}}\log\left(\frac{\log K}{\epsilon}\right) \right) = O\left( \frac{1}{p_{\text{min}}}\left(\log\left(\frac{1}{\epsilon}\right)+\log\log\left(\frac{1}{p_{\text{min}}}\right)\right)\right)\;.
\end{align}
%{\color{blue} (It looks to me that this should be \begin{align}
%O\left( \frac{1}{p_{\text{min}}}\left(\log\left(\frac{1}{\epsilon}\right)+\log\log\left(\frac{1}{p_{\text{min}}}\right)\right)\right)
%\end{align}}
The coefficients $c_j$ can be computed in time $\text{poly}\left(K,M\log(1/\epsilon) \right)$ and have 1-norm
\begin{align}
\|c\|_1 \in O\left(\log\left(\frac{\log(1/\epsilon)}{p_{\min}}\right) \right).
\end{align}
Now note that 
\begin{align}
ae^{ip}+be^{-ip} &= (a+b)\cos p+(a-ib)\sin p,\nonumber \\ e^{\pm i \pi m(1-p)/2} &= e^{\pm im\pi/2}e^{\mp i\pi m p}, \nonumber
\end{align}
allowing $f(p)$ to be rewritten as 
\begin{align}
f(x) = \sum_{m=1}^M \left( b^{(1)}_m \cos \left(\pi m p/2\right) +b^{(2)}_m \sin \left(\pi m p/2\right) \right),
\end{align}
where the coefficients $b^{(1)}_m$ and $b^{(2)}_m$ can be computed in time $O(1)$ from the coefficients $c_m$.

%Also from the lemma, the 1-norm of the coefficients in the Fourier series is bounded by $\|a\|_1 = O(\log K)=O(\log\log(1/\epsilon)+\log(1/{p_{\min}}))$.
%
%Thus, in the end, we have an $\epsilon$-approximation of $\text{Tr}(\rho \log \rho)$ by a series of the form 
%\begin{align}
%\tilde{S}(\rho) =\sum_{k=1}^{M}\left(\sum_{j=1}^N\rho_j\left( c^{(1)}_k\cos(\rho_jt_k)+c^{(2)}_k\sin(\rho_jt_k) \right)\right)
%\end{align}
%with $M=O(\log(1/\epsilon)/{p_{\min}})$ and
%\begin{align}
%\|c\|_1=\sum_k\left(\|c^{(1)}_k\|+\|c^{(2)}_k\| \right) = O(\log\log(1/\epsilon)+\log\log(1/(1-p_{\min})))\;.
%\end{align}

\end{proof}
%
%\subsubsection{Finding the Fourier approximation}
%
%So far we showed that an efficient Fourier approximation to the logarithm exists. Now we concern ourselves with actually computing the approximation. 

%\subsection{Density matrix exponentiation}
%
%Low et al. provide an efficient method of simulating an evolution $e^{-i\rho t}$ given access to an oracle $G$ that prepares a purified state version of $\rho$, i.e.,
%\begin{align}
%G\ket{0}^{\otimes \log{N}}\rightarrow\sum_{j=1}^N \sqrt{\rho}_j\ket{j}\ket{\chi_j} \;.
%\end{align}

We note that series approximations to von Neumann entropy have also been considered by Gilyen and Li~\cite{GL2020}. However, their approximation requires a number of terms that scales linearly with Hilbert-space dimension. We forego this explicit dependence on Hilbert-space dimension by choosing instead to express the dependence in terms of the cut-off $p_{\text{min}}$, which may be larger than $1/N$.

\begin{figure}[h]
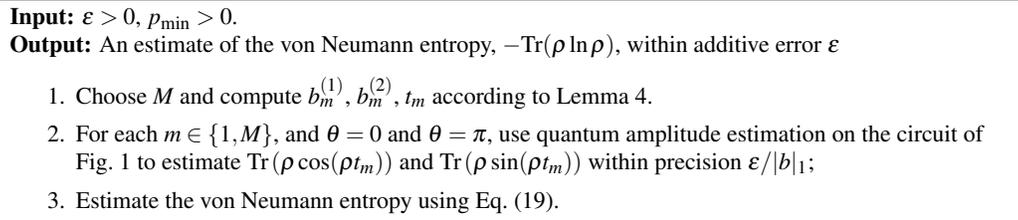
 \framebox{
	\begin{minipage}{.75\textwidth}
		\raggedright
		{\bf Input:} $\varepsilon>0$, $p_{\text{min}}>0$. \\
		{\bf Output:} An estimate of the von Neumann entropy, $-\text{Tr}(\rho\ln\rho)$, within additive error $\epsilon$  \\
		\begin{enumerate}
			\item Choose $M$ and compute $b^{(1)}_m$, $b^{(2)}_m$, $t_m$ according to Lemma \ref{lem:logfourierapprox}.
			\item For each $m\in\{1,M\}$, and $\theta=0$ and $\theta=\pi$, use quantum amplitude estimation on the circuit of Fig.~\ref{fig:entropyestcircuit} to estimate $ \text{Tr}\left(\rho\cos(\rho t_m)\right) $ and $ \text{Tr}\left(\rho\sin(\rho t_m)\right) $ within precision $ \epsilon/|b|_1 $;
			\item Estimate the von Neumann entropy using Eq.~\eqref{eq:entropy_approx}.
		\end{enumerate} 
	\end{minipage}}
	\caption{Entropy Estimation Algorithm}\label{alg:entropy_estimation}
\end{figure}

This immediately leads to an approximation to the von Neumann entropy as a linear combination of terms of the form $\text{Tr}(\rho \cos \rho t)$ and $\text{Tr}(\rho \sin \rho t)$ which is given by
\begin{align}\label{eq:entropy_approx}
    \tilde{S}(\rho) = \sum_{m=1}^{M}\left(b^{(1)}_m\text{Tr}(\rho\cos(p t_m))+b^{(2)}_m\text{Tr}(\rho sin(p t_m)) \right)\;,
\end{align}
where $b^{(1)}_m$, $b^{(2)}_m$, $t_m$ and $M$ are as in Lemma~\ref{lem:logfourierapprox}. The individual terms in the approximation can be estimated by repeated used of the quantum circuit in Fig.~\ref{fig:entropyestcircuit}. Lemma~\ref{lem:logfourierapprox} implies that
\begin{align}
    \left| \text{Tr}(-\rho\ln\rho)-\tilde{S}(\rho)  \right| \leq \varepsilon\;,
\end{align}
We provide an explicit algorithm in Fig.~\ref{alg:entropy_estimation} which uses Eq.~\eqref{eq:entropy_approx} to yield an estimate of the von Neumman entropy within a desired additive precision.

% \begin{enumerate}
% \item Choose $t_m$, $m \in \{1,M\}$ according to the conditions of Lemma \ref{lem:logfourierapprox}.
% \item For each $m\in\{1,M\}$, and $\theta=0$ and $\theta=\pi$, use quantum amplitude estimation on the circuit of Fig.\ \ref{fig:entropyestcircuit} to estimate the probabilities in equation \eqref{eq:probmeas} to precision $ \epsilon/|b|_1 $; this provides estimates of $ \text{Tr}\left(\rho\cos(\rho t_m)\right) $ and $ \text{Tr}\left(\rho\sin(\rho t_m)\right) $ within the same precision.
% \item Compute an estimate of the entropy as
% \begin{align} \label{eq:entropy_est_expr}
% \tilde{S(\rho)} = \sum_{m=1}^M \left(b_m^{(1)}\text{Tr}\left(\rho\cos(\rho t_m)\right)+b_m^{(2)}\text{Tr}\left(\rho\sin(\rho t_m)\right) \right),
% \end{align}
% where the coefficients are as defined in Lemma \ref{lem:logfourierapprox}.
% \end{enumerate}

%We choose $t_k$, $k\in{1,M}$ according to the conditions of Lemma \ref{lem:logfourierapprox}. Our algorithm uses quantum amplitude estimation on the circuit of Fig.\ \ref{fig:entropyestcircuit} to return an estimate of the probabilities as given in equation \eqref{eq:probmeas}. The entropy is estimated by summing for all $t_k$ with weights as given by Lemma \ref{lem:logtaylorapprox}.

\paragraph*{Complexity of entropy estimation}

We analyze the complexity of entropy estimation in terms of the uses of the oracle $ U_{\rho} $ which is expected to be the most expensive part of our algorithm. The evaluation also requires a number of two qubit gates to implement controlled operations, but these will at most be a constant factor more. 

A crucial component of our algorithm is the ability to simulate time evolution with $ \rho $, i.e., density matrix exponentiation given access to $ U_{\rho} $. An efficient protocol for achieving this was given by Low and Chuang \cite{lowchuang2016qubitization}. Their result implies that given access to a unitary $ U_{\rho} $ that prepares a purification $ \sum_j \sqrt{p_j} \ket{j}\ket{\psi_j} $ of a density matrix $ \rho = \sum_j p_j \ketbra{\psi_j}{\psi_j} $, time evolution with $ \rho $ for time $t$ and to precision $ \epsilon $ (in the spectral norm) can be simulated with a number of queries to $U_{\rho}$ that is in $ O(t+\log(1/\epsilon)) $.

We have chosen to evaluate each term in equation \eqref{eq:entropy_approx}
with the same precision $\epsilon'=\epsilon/\|b\|_1$. This is satisfied if we demand that our simulation be accurate to within precision $ \epsilon'/2 $, and that amplitude estimation return an estimate also with precision $ \epsilon'/2 $.
\begin{proofof}{Theorem~\ref{thm:est_entropy}}
Following Low and Chuang~\cite{lowchuang2016qubitization}, the cost of simulating time evolution with $\rho$ to precision $\epsilon'/2$ is 
\begin{align}
O\left(t+\log(1/\epsilon')\right).
\end{align}
The overall number of queries to $U_p$ needed to estimate a term in equation \eqref{eq:entropy_approx} using amplitude estimation (which requires exponentiation of $\rho$ for time $t$) within error $\epsilon$ and probability of failure at most $\epsilon$ is in,
\begin{align}
O\left(\left(\frac{\|b\|_1}{\epsilon}\right)\left(t+\log(1/\epsilon)+\log\|b\|_1\right) \right)
\end{align}

The exponentiation time $t_m$ takes values $m\pi$ for $m \in \left\{1,M\right\}$, where $ M $ is the maximum degree of the Fourier series approximation to the von Neumann entropy in Lemma \ref{lem:logfourierapprox}. Thus the overall cost of entropy estimation (in terms of the uses of $\rho$) turns out to be given by
\begin{align}
C_S & \in O\left(\left(\frac{\|b\|_1}{\epsilon}\right)\sum_{m=1}^M\left(m+\log(1/\epsilon)+\log\|b\|_1 \right) \right) \\
&= O\left(\left(\frac{\|b\|_1}{\epsilon}\right)\left(M^2+M\log(1/\epsilon)+M\log\|b\|_1 \right) \right) \\
%&= O\left(\frac{1}{\epsilon p_{\text{min}}^2}\log\left(\frac{1}{p_{\text{min}}}\right)\left(\log\left(\frac{1}{\epsilon}\right)+\log\log\left(\frac{1}{p_{\text{min}}}\right) \right)^2 \right) \\
&=\tilde{O}\left(\frac{1}{\epsilon p_{\text{min}}^2}\right). \label{eqn:entropyestcost}
\end{align}
\end{proofof}

\subsection{Energy Estimation}\label{sec:energy_estimation}

Estimating the energy for a mixed state, given an appropriate purification of the density operator, is a comparatively easier task than estimating the von Neumann entropy. The approach we take uses the Hadamard test in conjunction with ideas from linear-combinations-of-unitaries to estimate the average energy. However, other methods such as those of Knill et al.~\cite{knill2007optmeas} may also be employed for this problem. We assume that the Hamiltonian is represented as a linear combination of unitary operations,
\begin{align} \label{eq:Hamil_LCU}
H = \sum_{k=1}^K \alpha_k V_k
\end{align}
where the coefficients can be taken to be positive without any loss of generality. Efficient representations of such form exist for many systems of interest, e.g., qubit Hamiltonians. We further assume access to oracles $U_P$ and $U_S$ defined as follows:
\begin{align}
&U_P:~\ket{0}\longrightarrow \frac{1}{\sqrt{\left|\alpha\right|_1}}\sum_{k=1}^{K} \sqrt{\alpha_k}\ket{k}, \\
&U_S = \sum_{k=1}^K \ketbra{k}{k}\otimes V_k. 
\end{align}
We state below the performance of our algorithm in terms of these oracles.
\begin{theorem} \label{thm:est_avg_energy}
Let $H$ be a Hamiltonian that can be represented as in equation \eqref{eq:Hamil_LCU} for $K\in O(\log N)$, and let $U_{\psi}$ be a unitary that prepares a state $\ket{\psi}$. The number of applications of the unitaries $U_P$, $U_S$ and the unitary $U_{\psi}$ to estimate the average energy $\bra{\psi}H\ket{\psi}$ within error $\epsilon$ with probability at least $2/3$ is in
$O\left(\left\|\alpha\right\|_1/\epsilon\right)$.
\end{theorem}

\begin{figure} 
\begin{equation*}
\Qcircuit @C=1.5em @R=0.3 em {
               & \lstick{\ket{0}} & \gate{\rm H} & \qw & \ctrl{1} & \qw & \gate{\rm H} & \meter \\
               & \lstick{\ket{0}} & \qw {/}  & \gate{U_P} & \multigate{1}{U_S} & \gate{U_P^{\dagger}} & \qw  \\
               & \lstick{\ket{\psi}} & \qw {/}  & \qw & \ghost{U_S} & \qw & \qw
        }   
\end{equation*}
\caption{Quantum circuit for energy estimation}
\label{fig:energyestcircuit}
\end{figure}
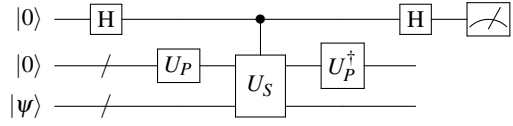

The average energy can be estimated from the measurement statistics of the first qubit in Fig.~\ref{fig:energyestcircuit}, as stated in the following lemma.
\begin{lemma} \label{lem:est_avg_energy}
The probability of obtaining outcome $ y = +1 $ from the circuit of Fig.\ \ref{fig:energyestcircuit} is
\begin{align}
\text{Pr}(+1) = \frac{1}{2}\left(1+\frac{\bra{\psi}H\ket{\psi}}{\left|\alpha\right|_1}\right)
\end{align}
\end{lemma}

\begin{proof}
	The state of the system in \ref{fig:energyestcircuit} before the final Hadamard is given by
	\begin{align}
	\ket{\Psi_f} = \frac{1}{\sqrt{2}}\left(\ket{0}\otimes\ket{0}_p\otimes\ket{\psi} + \ket{1}\otimes\frac{1}{\sqrt{\left|\alpha\right|_1}}\sum_{k=1}^K \sqrt{\alpha_k}U_P^{\dagger}\ket{k}_p\otimes V_k\ket{\psi}\right)
	\end{align}
	Therefore, the probability of obtaining outcome +1 on measuring the topmost qubit is given by,
	\begin{align}
	\text{Pr}(0)&=\text{Tr}\left(\ketbra{+}{+}\otimes\mathbb{I}_P\otimes\mathbb{I}\ketbra{\Psi_f}{\Psi_f} \right) =\text{Tr}\left(\braket{+}{\Psi_f}\braket{\Psi_f}{+}\right) \nonumber \\
	&=\frac{1}{4}\text{Tr}\Bigg(\ketbra{0}{0}_p\otimes\ketbra{\psi}{\psi} + \frac{1}{\left|\alpha\right|_1}\sum_{j,k=1}^K\sqrt{\alpha_j\alpha_k}U_P^{\dagger}\ketbra{j}{k}_pU_P\otimes V_j\ketbra{\psi}{\psi}V_k^{\dagger}\nonumber \\
	&\hspace*{4em} + \frac{1}{\sqrt{\left|\alpha\right|_1}}\sum_{k=1}^K\sqrt{\alpha_k}\left(\ketbra{0}{k}_pU_P\otimes\ketbra{\psi}{\psi}V_k^{\dagger} + U_P^{\dagger}\ketbra{k}{0}_p\otimes V_k\ketbra{\psi}{\psi}\right)\Bigg) \nonumber \\
	&=\frac{1}{4}\Bigg(1+\frac{1}{\left|\alpha\right|_1}\sum_{j,k=1}^K\sqrt{\alpha_j\alpha_k}\braket{j}{k}\braket{\psi}{\psi}\nonumber \\&\hspace*{4em}+\frac{1}{\sqrt{\left|\alpha\right|_1}}\sum_{k=1}^K\sqrt{\alpha_k}\left(\bra{k}U_P\ket{0}_P\bra{\psi}V_k^{\dagger}\ket{\psi}+\bra{0}U_P^{\dagger}\ket{k}_P\bra{\psi}V_k\ket{\psi}\right)\Bigg) \nonumber \\
	&=\frac{1}{2}\left(1+\frac{1}{\left|\alpha\right|_1}\sum_{k=1}^K\alpha_k\bra{\psi}V\ket{\psi}\right) \nonumber\\
	&=\frac{1}{2}\left(1+\frac{\bra{\psi}H\ket{\psi}}{\left|\alpha\right|_1}\right)
	\end{align}
	where we have used that $V^{\dagger}=V$ which is required for $H$ to be Hermitian. 
\end{proof}

\begin{proofof}{Theorem~\ref{thm:est_avg_energy}}
The proof follows constructively using the circuit laid out in Fig.~\ref{fig:energyestcircuit}.  As noted by Lemma~\ref{lem:est_avg_energy}, there exists a circuit that applies $O(1)$ queries to $U_s$ and $U_p$ such that the probability of observing the state in a marked sub-space (denoted by outcome-zero in the top-most qubit in Fig.~\ref{fig:energyestcircuit}).  As the circuit is unitary (if the top-most measurement is omitted) it satisfies the requirements of Theorem 12 of Brassard et al.~\cite{brassard2002quantum}, which implies that $O(1/\delta)$ queries are needed to learn $Pr(+1)=\frac{1}{2}\left( 1 + \frac{\bra{\psi}H \ket{\psi}}{|\alpha|_1}\right)$ within error $\delta$ with probability greater than $2/3$.  Thus if we wish to learn $\bra{\psi} H \ket{\psi}$ within error $\epsilon$ it suffices to take $\delta \le \epsilon/|\alpha|_1$.  This completes the proof.
\end{proofof}

\section{Free Energy Optimization}\label{sec:free_energy_optimization}
%\textbf{TODO: Emphasize that this is for theoretical purposes.  Discuss high-order or why we only want fixed order.}
There are many optimization methods that can be used in order to minimize the free energy.  In practice, gradient-free optimizers such as Powell's method are likely to be among the best approaches for finding the Gibbs state given stochastic noise in the entropy and energy evaluations.  However, gradient descent optimization has a notable advantage: the complexity of the optimization process is well understood in the context of strongly convex functions.  In this section we discuss the complexity of the overall optimization problem in such circumstances and show that if free energy is sufficiently smooth and if we assume strong convexity (or assume that all local optima are global optima) then  free-energy optimization using gradient-descent optimization is efficient.

We estimate the gradient using a first order difference formula, i.e., 
\begin{align}
\frac{\partial f(x)}{\partial x} = \frac{f(x+\delta)-f(x)}{\delta}+ O\left(\delta \max \left|\frac{\partial^2 f(x)}{\partial x^2}\right| \right)
\end{align}
The following lemma provides the complexity of computing the gradient in the free energy. 

\begin{lemma}{\label{lem:est_grad}}
Let $\rho: \mathbb{R}^N \mapsto \mathbb{C}^{D\times D}$ be Hermitian, full-rank, twice differentiable, $\|\rho^{-1}\| \le p_{\min}$, $\beta>0$ and let $H=\sum_{j} \alpha_j V_j$ be a Hermitian matrix where $V_j \in \mathbb{C}^{D\times D}$ is unitary and $\alpha_j>0$. 
A vector $G\in \mathbb{R}^N$ can be computed such that
$$
\| G-\nabla\left({\rm Tr}\left(\rho H \right)- \beta^{-1} {\rm Tr}(\rho \log(\rho))\right)\|\le \epsilon,
$$
using a number of queries that is in
$$
\widetilde{O}\left(\frac{N}{\epsilon^2}\left(\frac{\beta^{-1}}{p_{\min}^2} + \|\alpha\|_1\right)\left( \beta\|H\|{\rm Tr}\left( \left|\frac{\partial^2 \rho}{\partial \theta^2} \right|\right) +{\rm Tr}\left(\left|\rho\frac{\partial \rho^{-1}}{\partial \theta} \right| \right)^2 \right) \right)
$$
\end{lemma}
\begin{proof}

The function we are optimizing is 
\begin{align}
F(\theta) &= {\rm Tr}(\rho(\theta) H) - \beta^{-1}{\rm Tr}(\rho(\theta) \log(\rho(\theta))
\end{align}
We assume that the state $\rho(\theta)$ is twice differentiable. 
Now let $\partial_\theta^2 \rho(\theta) = Q^{-1}(\theta) D(\theta) Q(\theta)$ be the diagonalized form of the operator for diagonal matrix $D(\theta)$ and basis transformations $Q$.  We then have from Von-Neumann's inequality and the cyclic property of the trace operation that

\begin{align}
{\rm Tr}\left(\frac{\partial^2 \rho(\theta)}{\partial^2 \theta} H \right) &= {\rm Tr}\left(D(\theta) Q^{-1}(\theta) H Q(\theta) \right)\nonumber\\
&\le \sum_i D_{ii}(\theta) \lambda\left(Q^{-1}(\theta) H Q(\theta)\right)_{ii}.
\end{align}
Here $\lambda(\cdot)$ gives the diagonal eigenvalue matrix for the operator.  We then have from the triangle inequality that
\begin{align}
{\rm Tr}\left(\frac{\partial^2 \rho(\theta)}{\partial^2 \theta} H \right) &= {\rm Tr}\left(D(\theta) Q^{-1}(\theta) H Q(\theta) \right)\nonumber\\
&\le \sum_i |D_{ii}(\theta)| |\lambda\left(Q^{-1}(\theta) H Q(\theta)\right)_{ii}|\nonumber\\
&\le {\rm Tr}\left(\sqrt{\left(\frac{\partial^2\rho(\theta)}{\partial^2 \theta}\right)\left(\frac{\partial^2\rho(\theta)}{\partial^2 \theta}\right)^\dagger} \right)\|H\|\nonumber\\
& = {\rm Tr} \left(\left|\frac{\partial^2 \rho}{ \partial \theta^2} \right| \right) \|H\|.
\end{align}

Now we have the comparably more difficult task of computing an upper bound on the second derivative of the von Neumann entropy.  
\begin{equation}
\frac{\partial^2 {\rm Tr}(\rho \log \rho)}{\partial \theta^2} = {\rm Tr}\left( \frac{\partial^2 \rho}{\partial \theta^2} \log(\rho)\right) + 2 {\rm Tr}\left( \frac{\partial \rho}{\partial \theta} \frac{\partial \log(\rho)}{\partial \theta} \right) + {\rm Tr} \left( \rho \frac{\partial^2 \log(\rho)}{\partial \theta^2} \right).
\end{equation}

First, we have again from Von Neumann's trace inequality and the fact that the eigenvalues of $\rho$ are in $[p_{\min},1]$ that
\begin{equation}
 \left|{\rm Tr}\left( \frac{\partial^2 \rho}{\partial \theta^2} \log(\rho)\right) \right|\le \|\log(\rho^{-1})\| {\rm Tr}\left( \left|\frac{\partial^2 \rho}{\partial \theta^2} \right|\right)\le \log\left(\frac{1}{p_{\min}}\right) {\rm Tr}\left( \left|\frac{\partial^2 \rho}{\partial \theta^2} \right|\right).\label{eq:logmderiv}
\end{equation}

The remaining terms in~\eqref{eq:logmderiv} require the use of an integral expression for the derivative of the logarithm 
\begin{equation}
\frac{\partial }{\partial \theta} \log(\rho) = \int_0^1 (1 - t (1- \rho))^{-1} \frac{\partial \rho}{\partial \theta}(1 - t (1- \rho))^{-1}\mathrm{d}t := \int_0^1 A^{-1} \frac{\partial \rho}{\partial \theta}A^{-1}\mathrm{d}t.\label{eq:logmderiv2}
\end{equation}

%Next  let $\ket{\phi_j}$ be a complete orthonormal set of right eigenvectors of $\partial_\theta \rho$ with eigenvalues $\lambda_j$ then~\eqref{eq:logmderiv} yields
%\begin{align}
%{\rm Tr}\left( \frac{\partial \rho}{\partial \theta} \frac{\partial \log(\rho)}{\partial \theta} \right) &={\rm Tr}\left( \int_0^1\frac{\partial \rho}{\partial \theta} A^{-1} \frac{\partial \rho}{\partial \theta}A^{-1}\mathrm{d}t \right)\nonumber\\
% &= \sum_{j,k} \int_0^1\bra{\phi_j}\frac{\partial \rho}{\partial \theta} A^{-1} \ketbra{\phi_k}{\phi_k}\frac{\partial \rho}{\partial \theta}A^{-1}\mathrm{d}t \ket{\phi_j}.\nonumber\\
%&\le \left(\int_0^1 \|(1-t(1-\rho))^{-1}\| \mathrm{d}t \right)^2\sum_{j,k} |\lambda_j| |\lambda_k|\le  \max_{t} \|A^{-1}\|^2 {\rm Tr}\left(\left|\frac{\partial \rho}{\partial \theta} \right| \right)^2\nonumber\\
%&\le \frac{\log^2\left(\frac{1}{p_{\min}}\right){\rm Tr}\left(\left|\frac{\partial \rho}{\partial \theta} \right| \right)^2}{(1-p_{\min})^2}.\label{eq:2termderiv}
%\end{align}
%
%The final term in the second derivative of the von Neummann entropy can be computed using the fact that for any full-rank matrix $A$
%
%
%\begin{align}
%\partial_{\theta} 1 = \partial_{\theta}( A A^{-1})=0 &\Rightarrow (\partial_{\theta} A)A^{-1} = -A (\partial_{\theta} A^{-1})\nonumber\\
%&\Rightarrow \partial_{\theta} A^{-1} = -A^{-1} \frac{\partial A}{\partial \theta}A^{-1}.
%\end{align}
Next  let $\bra{\phi_j}$ be a complete orthonormal set of left eigenvectors of $\rho\partial_\theta \rho^{-1}$ with eigenvalues $\lambda_j$ then under the assumption that $\rho \partial_\theta \rho^{-1}$ is non-singular~\eqref{eq:logmderiv2} yields
\begin{align}
{\rm Tr}\left( \frac{\partial \rho}{\partial \theta} \frac{\partial \log(\rho)}{\partial \theta} \right) &={\rm Tr}\left( \int_0^1\frac{\partial \rho}{\partial \theta} A^{-1} \frac{\partial \rho}{\partial \theta}A^{-1}\mathrm{d}t \right)\nonumber\\
 &= \sum_{j,k} \int_0^1\bra{\phi_j}\frac{\partial \rho}{\partial \theta} A^{-1} \ketbra{\phi_k}{\phi_k}\frac{\partial \rho}{\partial \theta}A^{-1}\ket{\phi_j}\mathrm{d}t .\nonumber\\
 &= \sum_{j,k} \int_0^1\bra{\phi_j}\rho\frac{\partial \rho^{-1}}{\partial \theta}\rho A^{-1} \ketbra{\phi_k}{\phi_k}\rho\frac{\partial \rho^{-1}}{\partial \theta} \rho A^{-1} \ket{\phi_j}\mathrm{d}t.\nonumber\\
 &= \sum_{j,k} \int_0^1 \lambda_j \lambda_k\bra{\phi_j}\rho A^{-1} \ketbra{\phi_k}{\phi_k} \rho A^{-1}\ket{\phi_j}\mathrm{d}t .\nonumber\\
&\le  \int_0^1 \|\rho A^{-1}\|^2 \mathrm{d}t {\rm Tr}\left(\left|\rho \frac{\partial \rho^{-1}}{\partial \theta} \right| \right)^2\nonumber\\
&\le \max_{p\in [0,1]} \left( \int_0^1 \left|\frac{p}{1-t(1-p)} \right|^2 \mathrm{d}t \right)
{\rm Tr}\left(\left|\rho \frac{\partial \rho^{-1}}{\partial \theta} \right| \right)^2\nonumber\\
&=\max_{p\in [0,1]} \left( p {\rm Tr}\left(\left|\rho\frac{\partial \rho^{-1}}{\partial \theta} \right| \right)^2\right)={\rm Tr}\left(\left|\rho\frac{\partial \rho^{-1}}{\partial \theta} \right| \right)^2.\label{eq:2termderiv}
\end{align}

For our choice of $A$ in~\eqref{eq:logmderiv2} we specifically have that
\begin{equation}
 -A^{-1} \frac{\partial A}{\partial \theta}A^{-1}= -tA^{-1} \frac{\partial \rho}{\partial \theta} A^{-1}\label{eq:Aderiv}
\end{equation}
Using this we then have that
\begin{align}
{\rm Tr}\left(\rho\frac{\partial^2 \log(\rho)}{\partial \theta^2}\right) = \int_0^1 \rho A^{-1} \frac{\partial^2 \rho}{\partial \theta^2}A^{-1}+{\rm Tr} \left(\rho \frac{\partial A^{-1}}{\partial \theta}   \frac{\partial \rho}{\partial \theta} A^{-1}\right)+{\rm Tr} \left(\rho  A^{-1}\ \frac{\partial \rho}{\partial \theta}\frac{\partial A^{-1}}{\partial \theta} \right)\mathrm{d}t 
\end{align}
Now using the von Neumann trace inequality, the triangle inequality and the sub-multiplicative property of the spectral norm we have that
\begin{align}
\left|\int_0^1 {\rm Tr}\left(\rho A^{-1}\frac{\partial^2 \rho}{\partial \theta^2}A^{-1}\right)\mathrm{d}t  \right| &\le {\rm Tr}\left(\left|\frac{\partial^2 \rho}{\partial \theta^2} \right| \right)\int_0^1 \|A^{-1} \rho A^{-1}\|\mathrm{d}t \nonumber\\
&\le {\rm Tr}\left(\left|\frac{\partial^2 \rho}{\partial \theta^2} \right| \right)\max_{p\in [0,1]}\left(\int_0^1\frac{p}{(1-t(1-p))^2} \mathrm{d}t\right) \nonumber\\
&=  {\rm Tr}\left(\left|\frac{\partial^2 \rho}{\partial \theta^2} \right| \right).
\end{align}

Next we have from~\eqref{eq:Aderiv},~\eqref{eq:2termderiv} and the fact that $(\partial_\theta \rho A^{-1})^2$ is positive semi-definite that
\begin{align}
\left| {\rm Tr}\left(\int_0^1 \rho \frac{\partial A^{-1}}{\partial \theta}   \frac{\partial \rho}{\partial \theta} A^{-1}\right) \mathrm{d}t \right| &= \left| {\rm Tr}\left(\int_0^1 \rho t A^{-1}    \frac{\partial \rho}{\partial \theta} A^{-1} \frac{\partial \rho}{\partial \theta} A^{-1}\right) \mathrm{d}t \right|\nonumber\\
&\le {\int_0^1{\rm Tr}\left( \left|\left(\frac{\partial \rho}{\partial \theta} A^{-1} \right)^2 \right| \right)\mathrm{d}t} \max_t\|\rho A^{-1}\|\nonumber\\
&\le {\rm Tr}\left(\left|\rho \frac{\partial \rho^{-1}}{\partial \theta} \right| \right)^2.
\end{align}
By following the exact same argument on the remaining mixed derivative of this form, we then have 
\begin{equation}
\left|{\rm Tr}\left(\rho\frac{\partial^2 \log(\rho)}{\partial \theta^2}\right)\right|  \le {\rm Tr}\left(\left|\frac{\partial^2 \rho}{\partial \theta^2} \right| \right)+ 2{\rm Tr}\left(\left|\rho\frac{\partial \rho^{-1}}{\partial \theta} \right| \right)^2
\end{equation}
Finally we then have that
\begin{equation}
\frac{\partial^2 {\rm Tr}(\rho \log \rho)}{\partial \theta^2}\in O\left( \log\left(\frac{1}{p_{\min}}\right) {\rm Tr}\left( \left|\frac{\partial^2 \rho}{\partial \theta^2} \right|\right) +{\rm Tr}\left(\left|\rho\frac{\partial \rho^{-1}}{\partial \theta} \right| \right)^2 \right)
\end{equation}
and
\begin{align}
&\delta \frac{\partial^2F}{\partial \theta^2}\in O\left(\beta^{-1}\delta(\log(p_{\min}^{-1}) + \beta\|H\|){\rm Tr}\left( \left|\frac{\partial^2 \rho}{\partial \theta^2} \right|\right) +{\rm Tr}\left(\left|\rho\frac{\partial \rho^{-1}}{\partial \theta} \right| \right)^2 \right)
\end{align}

It then follows from the triangle inequality that by combining $N$ such gradient evaluations we can estimate the gradient of the free energy using a step size of
\begin{align}
\delta \in \Theta\left(\frac{\epsilon\beta}{{N}}\left( {(\log(p_{\min}^{-1}) + \beta\|H\|){\rm Tr}\left( \left|\frac{\partial^2 \rho}{\partial \theta^2} \right|\right) +{\rm Tr}\left(\left|\rho\frac{\partial \rho^{-1}}{\partial \theta} \right| \right)^2} \right)^{-1} \right)\;.
\end{align}
Also, each of the evaluations of $F(\theta)$ must be carried out to precision $\epsilon\delta$ which implies that the entropy must be estimated within error $\beta \epsilon \delta$ and from Theorems~\ref{thm:est_entropy} and~\ref{thm:est_avg_energy} the number of queries needed to estimate the gradient of the free energy is in

\begin{equation}
\widetilde{O}\left(\left(\frac{\beta^{-1}}{p_{\min}^2} + \|\alpha\|_1\right)\frac{\beta}{\epsilon\delta} \right)\subseteq \widetilde{O}\left(\frac{N}{\epsilon^2}\left(\frac{\beta^{-1}}{p_{\min}^2} + \|\alpha\|_1\right)\left( \beta\|H\|{\rm Tr}\left( \left|\frac{\partial^2 \rho}{\partial \theta^2} \right|\right) +{\rm Tr}\left(\left|\rho\frac{\partial \rho^{-1}}{\partial \theta} \right| \right)^2 \right) \right)
\end{equation}
\end{proof}
%\section{Convergence of Gradient Descent}

The above analysis shows that the gradient can be computed using our method using a relatively small number of queries.  While the cost per gradient evaluation is not necessary prohibitive, a challenge remains in estimating the number of steps required to obtain convergence.  The following provides an estimate of the number of steps required to get close to the minimum free energy, neglecting errors in the free energy estimation.

The flavor of gradient descent that we consider has the following form for  a function $f(x)$,
\begin{align}
x_{t+1}=x_t-r\nabla f(x_t)
\end{align}
where $r$ is the rate of descent.
There are bounds on the convergence of gradient descent in the scenario where the function to be optimized is smooth and strongly convex.
$f(x)$ is said to be $\mu$-strongly convex if
\begin{align}
f(x')\geq f(x) +\nabla f(x)\cdot(x'-x)+\frac{\mu}{2}\|x-x'\|^2\;.
\end{align}
$f(x)$ is said to be $L$-smooth if
\begin{align}
f(x')\leq f(x) +\nabla f(x)\cdot(x'-x)+\frac{L}{2}\|x-x'\|^2\;.
\end{align}
For a function to be both smooth and strongly convex means that it is both upper and lower bounded by quadratic functions.  Even if a function is not strongly convex, then the objective function can be made strongly convex within the vicinity of an optima by adding a quadratic function to the objective function that has negligible impact on the derivatives.  Thus the assumptions of strong convexity can approximately hold locally even for landscapes that are littered with local optima.  Under such assumptions, the following theorem holds~\cite{Boyd}:
\begin{theorem}
Suppose that $f$ is both $\mu$-strongly convex and $L$-smooth, and has a minimum at $x=x^*$ Then, for a rate of descent $r=1/L$, 
\begin{align}
\|x(t)-x^*\| \leq \left(1-\frac{\mu}{L}\right)^t\|x(0)-x^*\|\;,
\end{align}
i.e., it converges to within $\epsilon$ distance of $x^*$ in time  $O\left((L/\mu)\log\left(\|x_0-x^*\|/\epsilon\right)\right)$
\end{theorem}

This result shows that at most a logarithmic number of repetitions will be needed in order to follow the gradients to the global optima for a strongly convex function.  Thus for constant $\mu$, $L$ and $x_0$ the complexity of gradient descent varies from the complexity of gradient evaluation only up to logarithmic factors.  Thus the results of Lemma \ref{lem:est_grad} also give the query complexity of performing the gradient descent optimization under such assumptions.

\section{Near term applications}\label{sec:near_term}

The current state-of-the-art quantum algorithms for Gibbs state preparation are desgined for fault-tolerant devices and in particular require significant circuit depth and a large overhead in terms of ancilla qubits, rendering them unsuitable for use in near-term quantum devices. The variational approach can be beneficial as it opens the possibility of discovering algorithms for preparing Gibbs states of systems of interest with much shorter depth. However, as with other quantum circuit optimization/learning approaches, there is always a significant possibility of the optimizer getting lost in Hilbert space and thus not converging in any reasonable time. Such pitfalls can be avoided by choosing a suitable ansatz or template for the algorithm, thereby restricting the search space and increasing the confidence that the optimizer converges to a circuit that is good enough, if not necessarily optimal. 

%We provide a proof of principle demonstration of the validity of this approach. We consider as our ansatz a perturbation about a trotterized adiabatic evolution. Suppose that we start in the coherent Gibbs state of a diagonal Hamiltonian $H_0$, and evolve with a time-dependent Hamiltonian that smoothly varies from $H_0$ to $H$, the latter being the Hamiltonian whose Gibbs state we are interested in. Provided that the Hamiltonian stays gapped, and that the evolution is sufficiently slow, this procedure will prepare the coherent Gibbs state of $H$. A trotterized digital simulation of this evolution would in turn prepare a good approximation to the Gibbs state with high probability. A way to validate our variational protocol is to therefore check whether it can find this trotterized adiabatic evolution given an initial ansatz that is essentially a perturbation about it. 
%
%In our simulations we take initial Hamiltonians of the form 
%\begin{align}
%H_0 = \sum_{j=1}^n a_j \sigma^{(j)}_Z
%\end{align}
%and investigate preparing Gibbs states for Hamiltonians $H = H_0+H_1$, where 
%\begin{align}
%H_1 = \sum_{j} b_j \sigma^{(j)}_X + \sum_{j,k} c_{j,k}\sigma^{(j)}_Z\sigma^{(j)}_Z.
%\end{align}

There are many ansatzae that could be considered in our variational approach.  We focus here on Trotterized adiabatic state preparation, which chooses an ansatz that strongly resembles a Trotter series approximation to a state preparation protocol for the eigenstates of the Hamiltonian $H$ from the eigenstates $H_0$ which are trivial to prepare.  Specifically, we define
\begin{equation} \label{eq:Hamil_form}
H'(s) = H_0 + sH_1 
\end{equation}
with $ H=H'(1)=H_0+H_1 $
Here $H'(1) = H$ and $H'(0)=H_0$, which means that if $H'(s)$ is gapped for all $s\in [0,1]$ then an adiabatic sweep along a linear adiabatic path from $H'(0)$ to $H'(1)$ will map the eigenstates of $H_0$ to those of $H=H_0+H_1$.
%Let for all $H$, $H(s) = \sum_{j=1}^m h_j(s) H_j$ then we take our state to be of the form for $\theta = [\theta_1,\ldots,\theta_r, \theta_{\rm prep}]$
%\begin{equation} \label{eq:ansatz}
%\ket{\psi_0(\theta)} = \prod_{k=1}^{r} e^{-i H([k-1]/r) T/r} U({\theta_{\rm prep}}) \ket{0} = \prod_{k=1}^{r} \prod_{j=1}^m e^{-i [\theta_k]_jH_j) T/r}\sum_j a_j(\theta_{\rm prep}) \ket{j}\ket{j},
%\end{equation}
We take our state to be of the form for parameters $\vec{\theta} = [\theta_1,\ldots,\theta_r, p_1,\ldots,p_{D-1}]$
\begin{align} \label{eq:ansatz}
\ket{\psi_0(\vec{\theta})} &= \prod_{k=0}^{r+1} e^{-i H'((\theta_k+\theta_{k+1})/2)\cdot(\theta_{k+1}-\theta_k) T} U(p_1,\ldots,p_{D-1}) \ket{0} \nonumber \\
&= \prod_{k=1}^{r} e^{-i H'((\theta_k+\theta_{k+1})/2)\cdot(\theta_{k+1}-\theta_k) T}\sum_{j=1}^D \sqrt{p_j} \ket{j}\ket{j},
\end{align}
where $ \theta_0=0 $ and $ \theta_1=1 $ and $ p_D =1-\sum_{j=1}^{D-1}p_j $ by normalization. In other words, the parameters $ \theta_1,\ldots,\theta_r $ specify the adiabatic path whereas $ p_1,\ldots,p_{D-1} $ are the probabilities of the different eigenstates in the density matrix obtained after tracing out one subsystem.
While the total time of evolution $T$ could additionally be varied over we take it to be fixed in the optimization for simplicity and the fact that $T$ is linearly dependent on the other parameters of the ansatz. 
This ansatz has one key advantage: if we take $T\rightarrow \infty$ and $r\rightarrow \infty$ then all Gibbs states can be prepared in this fashion.  This is to say, we know that a solution exists for the preparation of the true Gibbs state that is of this form.  Other solutions, which may require a sub-exponential number of parameters, may be preferable in practice to ours but we choose this ansatz strictly because we know that it will converge under the assumption that the gap of the Hamiltonian is bounded.  

\begin{figure}
\begin{minipage}[c][]{\textwidth}
	\centering
	(a)	\includegraphics[width=0.75\textwidth]{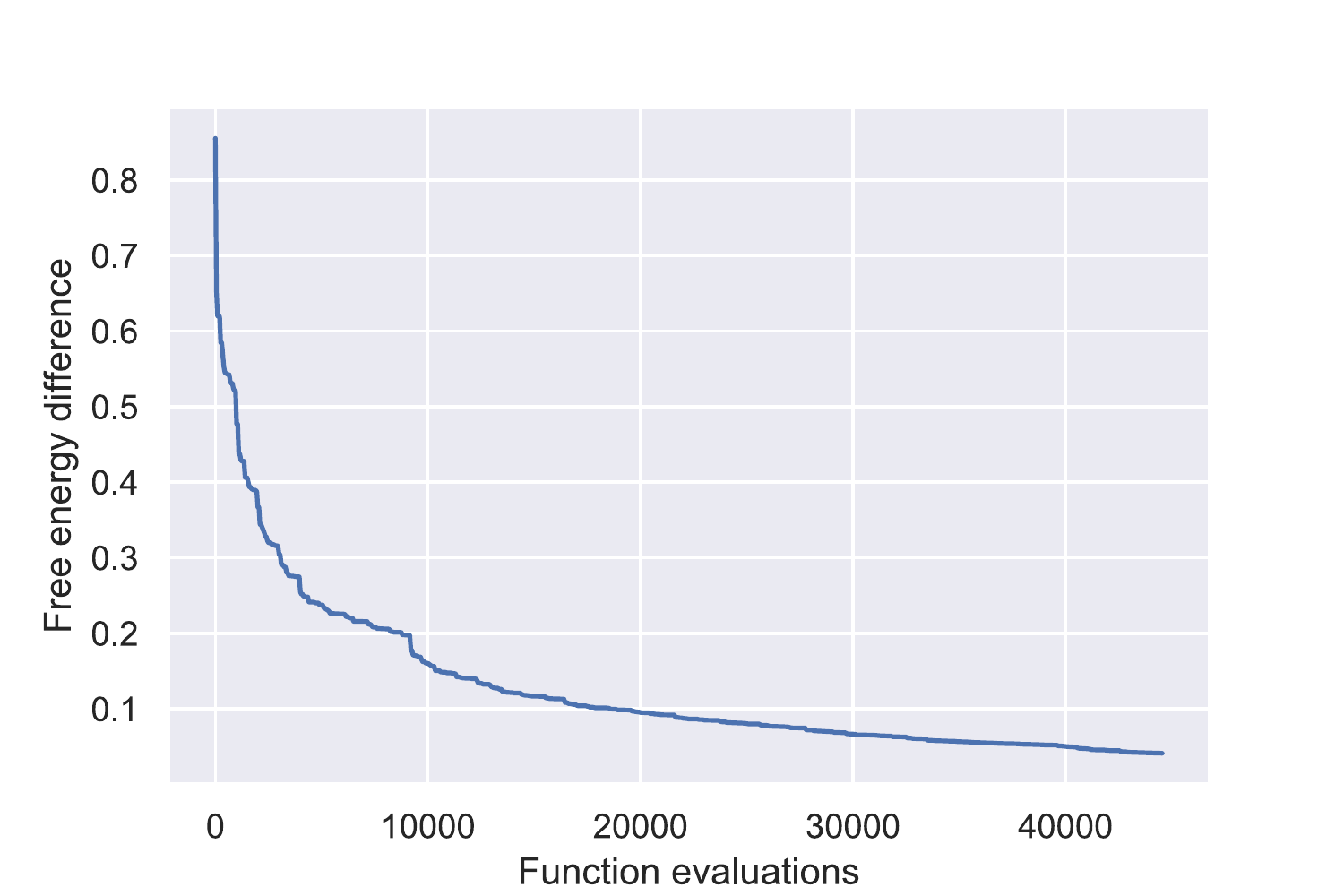}
	\par
	(b) \includegraphics[width=0.75	\textwidth]{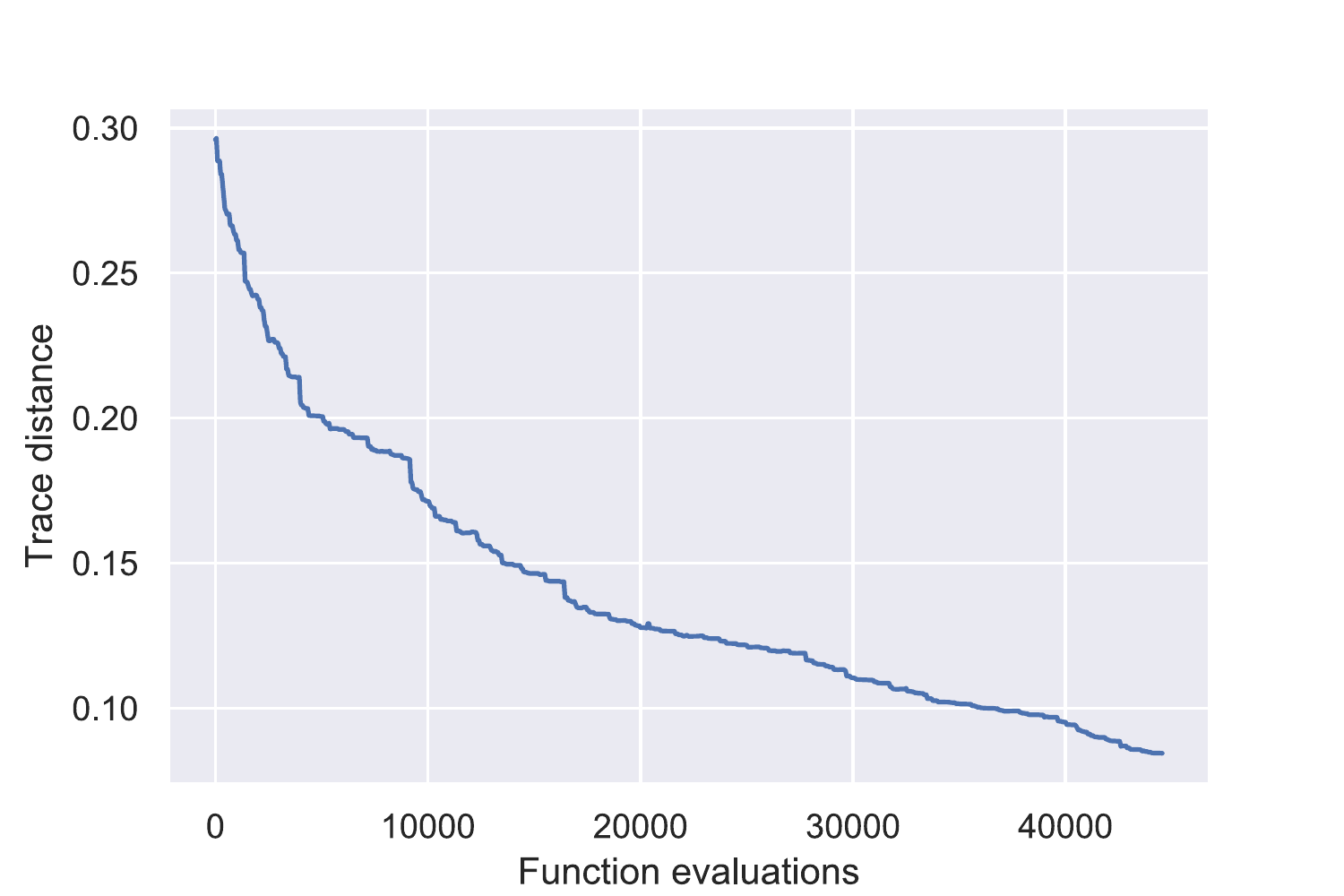}
\end{minipage}
\caption{Plots showing convergence to the Gibbs state using our variational ansatz for a five qubit Hamiltonian. 
(a) shows the difference between the lowest free energy reported by the optimizer up to the corresponding number of function evaluations and the free energy of the Gibbs state. (b) shows the trace distance between the true Gibbs state and the state corresponding to the lowest free energy obtained by the optimization. We see convergence in both the free energy and trace distance.}
\label{fig:convergence1}
\end{figure}

\subsection{Numerical Simulations}

We aim to demonstrate the validity of our protocol by numerically minimizing the free energy for the above ansatz. We added a penalty term to the free energy in the cost function to ensure that the parameters $ p_1,\ldots,p_D $ are valid probabilities, i.e., they are between 0 and 1. The variables $ \theta_1,\ldots,\theta_r $ were further parameterized as $ \theta_1=\tanh{\phi_1},\ldots,\theta_r = \tanh{\phi_{r}} $ to ensure that the total time of evolution remains bounded. We numerically minimized the cost function using Powell's conjugate direction method which is a gradient-free numerical optimization algorithm.

For our simulations, we restrict ourselves to qubit Hamiltonians of of the following form:
\begin{align}
H_0 &= \sum_{j=1}^n a_j \sigma^{(j)}_Z,\\
H_1 &= \sum_{j} b_j \sigma^{(j)}_X + \sum_{j,k} c_{j,k}\sigma^{(j)}_Z\sigma^{(j)}_Z,
\end{align}
the goal being to prepare Gibbs states of $ H=H_0+H_1 $. 
Fig.~\ref{fig:convergence1} shows the representative progress in numerical optimization for one such Hamiltonian with coefficients of magnitude $ O(1) $.
We chose an initialization such that the parameters $ p_1,\ldots,p_D $ were random perturbations about the true Gibbs state probabilities, and the path specified by the parameters $ \phi_1,\ldots,\phi_r $ was a random perturbation about a linear adiabatic path. We record the progress of the optimization as follows. At fixed intervals (of number of cost function evaluations), we have the optimizer return the quantum state at that point and the corresponding free energy. From this data, we compute the lowest free energy obtained till a given a number of function evaluations, and the corresponding quantum state. In fig.~\ref{fig:convergence1}(a) we plot the difference between this ``current best'' free energy and that of the true Gibbs state. Fig.~\ref{fig:convergence1}(b) plots the trace distance between the state corresponding to the ``current best'' free energy and the true Gibbs state. The results show that we indeed converge towards the minimum free energy in the optimization, and that this also implies convergence in trace distance to the true Gibbs state.

\section{Conclusion}
Gibbs state preparation remains an indispensible tool in quantum simulation and quantum machine learning.
While methods are known for preparing Gibbs states accurately on fault-tolerant quantum computers, existing approaches require complicated control structures or use heuristic methods derived from quantum thermodynamics.
Here we take a different approach that uses a linear combination of unitaries-based approach, that can be implemented semi-classically, to approximate the free energy.  Since the Gibbs state minimizes free energy
minimizing it is a natural target for such state preparation.  We show that high-temperature Gibbs states can be efficiently prepared using this approach and importantly, this approach allows the user to take advantage of prior knowledge about the form of the Gibbs state through a clever use of a variational ansatz.

We further show the method working, albeit slowly, for learning the Gibbs state for a five qubit system.  This work shows that the approach given here is likely to be viable on present day hardware and thus future work based on this could include the first training of quantum Boltzmann machines that do not use quantum annealing hardware.

Beyond this, a number of important problems remain open surroundding this work.  While our work showed that in principle minimizing the free energy is a valid objective function, if one wishes to minimize a statistical difference between the output state and the ideal Gibbs distribution then formally speaking it is not necessarily true that minimizing the free energy will minimize the distance if local optima are present in the landscape.  Addressing this issue as well as optimizing this approach to use state of the art optimizers and ansatzae that are more appropriate for the problem then trotterized adiabatic state preparation.  By finding such improved methods we may find ourselves one step closer to a practical understanding of how to train quantum Boltzmann machines and in turn bring us one step closer to realizing them in hardware.
\bibliographystyle{unsrt}
\bibliography{biblio}
\end{document}